# Améliorer les performances de l'industrie logicielle par une meilleure compréhension des besoins

Benjamin CHEVALLEREAU[1,3], Alain BERNARD[3], Pierre MÉVELLEC[2]

[1] IRCCyN
École Centrale de Nantes, 1 rue de la Noë, BP 92101, 44321 Nantes Cedex 3, France
{benjamin.chevallereau;alain.bernard}@irccyn.ec-nantes.fr

[2] IAE de Nantes
Chemin de la Censive du Tertre, BP 52231, 44322 NANTES Cedex 3
pierre.mevellec@univ-nantes.fr

[3] BlueXML
40 boulevard Jean Ingres, 44100 Nantes
bchevallereau@bluexml.com

*Résumé* - Les organisations actuelles se structurent et agissent en s'appuyant sur leurs systèmes d'information. Malgré les progrès considérables réalisés par la technologie informatique, on constate que les acteurs restent très souvent critiques par rapport à leur systèmes d'information. Une des causes de cet écart entre les espoirs et la réalité trouve sa source dans la difficulté à produire un cahier des charges suffisamment détaillé pour les opérationnels et interprétable par les spécialistes des systèmes d'information. Notre proposition vise à surmonter cet obstacle en organisant l'expression des besoins dans un langage commun aux opérationnels et aux experts techniques. Pour cela, le langage proposé pour exprimer les besoins est basé sur la notion de but. L'ingénierie dirigée par les modèles est présente à toute les étapes, c'est-à-dire au moment de la capture et de l'interprétation.

*Abstract* – Actual organization are structured and act with the help of their information systems. In spite of considerable progresses made by computer technology, we note that actors are very often critical on their information systems. Difficulties to product specifications enough detailed for functional profile and interpretable by information system expert is one of reason of this gap between hopes and reality. Our proposition wants to get over this obstacle by organizing user requirements in a common language of operational profile and technical expert.

*Mots clés* – ingénierie des besoins, objectif, modèle.

## 1 Introduction

Les systèmes informatiques sont de plus en plus complexes et nous aident à réaliser la plupart de nos activités. De plus, le volume d'information partagé, consulté ou stocké augmente continuellement en raison de l'informatisation généralisée de l'ensemble des systèmes. Nous devenons donc de plus en plus dépendant des différents systèmes d'information auxquels nous avons accès. Cette dépendance peut nous permettre d'améliorer significativement notre efficacité mais présente également de nombreux risques de perte de productivité dans le cas d'inadéquation entre le processus et l'outil le supportant. Au vu des différents rapports concernant l'industrie logicielle, il est réellement fondé d'étudier la phase de spécification des besoins et de s'intéresser à la compréhension des différents besoins afin de proposer de meilleures solutions répondant aux exigences du client.

La communication entre un expert technique et un opérationnel est extrêmement difficile. Ils n'utilisent pas la même sémantique. Ils ne connaissent pas le métier de l'autre et donc ne savent pas formuler clairement leurs idées sous une forme compréhensible. L'ingénierie des besoins a donc pour objectif d'améliorer cette communication afin de mieux comprendre les besoins des opérationnels et ainsi réaliser un outil complet et adapté. Pour résumer, elle a pour tâche de fournir une spécification des besoins qui doit être aussi complète et documentée que possible en utilisant des formats variés et adaptés de représentation afin d'établir une communication suffisante entre les différents participants impliqués[Pohl, 1996].

Le problème majeur, à l'heure actuelle, dans les méthodologies proposées par l'ingénierie des besoins est la complexité. Les différents formalismes restent très techniques et donc peu accessibles aux opérationnels. Le nombre de concepts proposés en font des méthodologies très riches mais, également, très difficiles d'accès en raison de leur complexité de compréhension. La simplicité du langage est dans notre proposition une caractéristique recherchée et indispensable. De plus, les méthodologies proposées par l'ingénierie des besoins sont généralement peu ou pas outillées. Il est donc impératif de fournir des outils accessibles afin de profiter pleinement de la simplicité du formalisme.

L'objectif d'amélioration de la communication peut également s'accompagner d'une amélioration de la productivité. L'expression des besoins doit jouer un rôle central dans le processus de développement. Pour cette raison, nous proposons à travers notre méthodologie un ensemble de transformations afin de traduire automatiquement l'expression des besoins. Cette méthodologie s'inscrit dans l'approche IDM[Kent, 2002] (« Ingénierie Dirigée par les Modèles ») qui en s'appuyant sur la notion de modèle vise à pérenniser les outils métier grâce à l'automatisation des processus réalisés auparavant à la main par les ingénieurs expérimentés. L'expression des besoins est vue comme un modèle. Ce dernier devient la source des différentes transformations proposées afin d'obtenir automatiquement les différentes interprétations désirées et nécessaires au passage du cahier des charges au système d'information.

Nous allons, tout d'abord, introduire dans la première partie l'état actuel de l'industrie logicielle et les réponses apportées par l'ingénierie des besoins. Nous pourrons ensuite aborder notre proposition de méthodologie en détaillant ses trois piliers qui sont un formalisme simple, un outil accessible et un ensemble de transformations automatiques. Pour conclure, nous nous intéresserons aux résultats obtenus et aux différentes pistes avancées.

## 2 CONTEXTE

### 2.1 État actuel de l'industrie logicielle

Traditionnellement, l'ingénierie des systèmes d'information se concentre sur la modélisation conceptuelle. Celle-ci vise à abstraire la spécification du système requis à partir de l'analyse des informations nécessaires à la communauté des utilisateurs. Cette spécification se concentre sur ce que doit faire le système, c'est-à-dire ses fonctionnalités[Rolland, 2007]. Bien que la modélisation conceptuelle permet aux profils techniques de comprendre la sémantique du domaine, elle ne permet pas de construire des systèmes acceptés par la communauté des utilisateurs. En effet, de nombreuses études [Robbins-Gioia LLC, 2002][The Standish Group, 1994][Cooke et al., 2001] montrent que les systèmes réalisés répondent de manière inadéquate ou insuffisante aux besoins exprimés par les utilisateurs. Le processus de développement actuel comporte donc des faiblesses. Celles-ci impliquent d'adapter les systèmes en cours ou après la réalisation, ce qui oblige à fournir une quantité d'effort très importante [Johnson, 1995].

Afin d'apporter des réponses adaptées à la communauté des utilisateurs, il est nécessaire de considérer les systèmes d'information comme un moyen d'atteindre un but déterminé dans une organisation. Comprendre ce but est une condition nécessaire à la conceptualisation d'un système. Il est important de dépasser la définition des fonctionnalités basée sur la vue du modèle conceptuel et d'étendre l'approche « que doit réaliser le système » vers « pourquoi le système est nécessaire ». La seconde question permet d'extraire les objectifs organisationnels et leur impact sur le système d'information.

À la vue des différents analyses [Krob, 2005], il est possible d'en extraire deux conclusions. Tout d'abord, il est évident qu'il est très difficile d'obtenir des mesures fiables de la performance de l'industrie logicielle [Glass, 2006]. La seconde conclusion nous montre qu'il existe un réel problème dans l'industrie logicielle et que la qualité des produits peut être améliorée. Après avoir identifié le problème à résoudre, d'autres analyses nous permettent d'identifier les principales raisons de ce taux d'échec. La définition des besoins et sa compréhension représentent les problèmes majeurs pour réaliser des produits répondant aux exigences des utilisateurs.

### 2.2 Ingénierie des besoins

L'objectif de l'ingénierie des besoins est de proposer des méthodologies, des formalismes et des outils pour apporter une réponse à cet ensemble de problèmes. Elle va permettre de mieux identifier, comprendre, valider et interpréter les besoins des utilisateurs afin de réaliser le système. Elle a été initialement décrite comme étant : « la partie du développement au cours de laquelle les gens tentent de découvrir ce qui est désiré »[Gause et Weinberg, 1989]. C'est pour cette raison que les premières méthodes se sont concentrées sur la définition du système donc ce qui devait être réalisé, en d'autres termes le « quoi ? ». Puis elles se sont concentrées sur comment le système devrait être défini, organisé puis implanté donc le « comment ? ». Ces différentes définitions restaient très techniques et inaccessibles à des opérationnels.

Afin d'impliquer les différents profils opérationnels, les approches se sont attachées aux raisons pour lesquelles le système est nécessaire, donc le « pourquoi ? ». Dans celles-ci, le système désiré est vu comme un moyen d'atteindre les différents objectifs de l'entreprise afin de répondre à ses problèmes organisationnels. La définition des besoins ne définit plus le système à réaliser mais les activités supportées par le système. Deux grandes familles ont été étudiées pour identifier les besoins. La première est centrée autour de la notion de scénario tandis que la seconde autour de la notion de but.

2.2.1 Ingénierie basée sur les scénarios (ou *Scenario-Based Requirements Engineering*)

Comme le décrit A. Sutcliffe[Sutcliffe, 2003], un scénario est une représentation du monde réelle. Dans le cas de l'ingénierie des besoins, un scénario est limité à la compréhension du système par les participants et se situe donc au niveau d'abstraction du système. Les événements sont donc des interactions entre l'utilisateur et le système. Il existe deux grandes méthodes proposées dans cette approche : la méthode ScenIC[Potts, 1999] et la méthode SCRAM[Sutcliffe et Ryan, 1998].

2.2.2 Ingénierie orientée par les buts (ou *Goal-Oriented Requirements Engineering*)

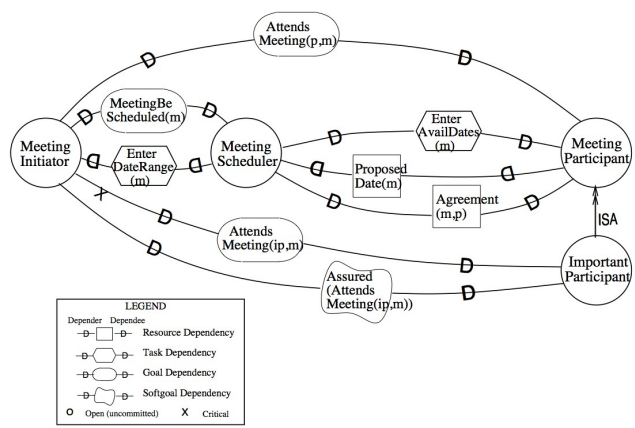

**Figure 1. Exemple avec le framework i***

Comme le définit A. Lamsweerde[Lamsweerde, 2001], un but doit être atteint par le système. Il peut donc être fonctionnel, c'est-à-dire qu'il représente un service fourni, ou non-fonctionnel, c'est-à-dire la qualité de service. Dans la majorité des cas, les buts sont généralement fixés implicitement par les participants. À partir du moment où l'analyse préliminaire des besoins a été faite, d'autres buts peuvent être obtenus par raffinement, abstraction...

2.2.3 Comparaison

Les buts se concentrent sur l'abstraction des besoins décrits de manière ambiguë par les utilisateurs. Les scénarios permettent de compléter ces buts en précisant comment le système va fonctionner pour répondre à la demande. Les techniques orientées par les buts sont généralement plus simples et requièrent moins d'efforts en comparaison des techniques basées sur les scénarios. La mise en place d'une méthodologie à l'aide des scénarios nécessite plus de temps et de ressources qu'une approche par les buts. Pour conclure, les scénarios permettent de préciser les buts[Misra et al., 2005].

Les méthodes proposées en ingénierie orientée par les buts (ou « GORE » en anglais pour *Goal-Oriented Requirements Engineering* sont généralement complexes à mettre en oeuvre et surtout à comprendre comme le montre l'exemple sur la figure 1 [Yu, 1997] qui reste illisible pour une personne non-experte dans cette méthodologie. Les difficultés de prise en main et de compréhension sont de véritables verrous à une bonne utilisation de ces méthodologies basées sur les buts.

*2.3 Organisation des flux de communication*

Afin de schématiser une organisation de flux de communication lors de la phase d'expression des besoins, nous définissons deux familles de participants. La première famille, identifiée sous le terme « expert système » regroupe l'ensemble des profils techniques rattachés généralement à une société de services qui a pour tâche de réaliser le système attendu. La seconde famille, identifiée par le terme « expert métier », regroupe l'ensemble des profils dits fonctionnels. Elle regroupe les futurs utilisateurs de l'application, les clients, les consultants. Pour résumer, elle englobe l'ensemble des profils possédant des connaissances sur le domaine d'activité supporté par le futur outil.

On peut schématiser une organisation traditionnelle de communication entre un expert système et un expert métier comme une interprétation puis une traduction de l'expression des besoins (voir figure2). L'expert système attend une spécification formelle, claire et précise définissant le système à réaliser pour initier le processus de développement. L'expert métier a un besoin, celui-ci est exprimé, dans le meilleur des cas, dans un cahier des charges ou dans un ensemble de documents décrivant l'activité à supporter. Cette expression du besoin est donc généralement rédigée à l'aide d'un langage naturel. Une étape de traduction est donc nécessaire à partir d'une définition informelle et ambigüe vers une spécification formelle. Cette étape est réalisée par un profil qui doit avoir les compétences d'un expert système afin de rédiger des spécifications utilisables mais également les compétences nécessaires à la compréhension du domaine d'activité de l'expert métier. De plus, ce profil doit connaître précisément l'environnement de cet expert métier : ses objectifs, ses ambitions, ses problèmes et son vocabulaire technique.

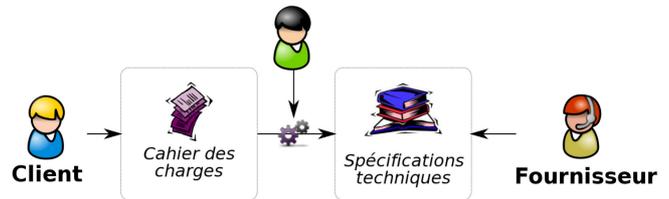

**Figure 2. Organisation traditionnelle**

L'étape d'interprétation puis de traduction est l'une des étapes la plus sensible du processus de développement logiciel. Si la définition des besoins est erronée ou incomplète avant le début du processus de développement, alors il existe une forte probabilité pour que le projet soit un échec partiel ou total. De plus, les méthodologies actuelles considèrent généralement que la spécification du besoin acquise à l'issue de la phase d'analyse du besoin représente un référentiel stable et fiable. Il est donc important de valider cette spécification par la majorité des acteurs de ce projet à l'aide de différents supports et moyens de communication. Notre proposition cherche à approfondir cette phase qui est probablement la plus importante dans le cycle du développement avec un nouveau langage et un mécanisme d'interprétation automatique.

## 3 PROPOSITION

*3.1 Motivation*

L'expression des besoins est la définition des exigences du futur système obtenu à partir des connaissances métier apportées par un ou plusieurs experts métier. Cependant, les « modèles métier », permettant de définir le domaine d'activité et les objectifs à atteindre, et les « modèles techniques », permettant de spécifier précisément le futur système à réaliser, ne sont pas exprimés dans les mêmes espaces sémantiques ce qui rend difficile la bonne coordination entre les différents profils. Les trois piliers de notre approche visent à surmonter cet obstacle en organisant l'expression des besoins dans un langage commun aux opérationnels et aux experts techniques et en fournissant un mécanisme d'interprétation permettant d'améliorer la communication entre les différents acteurs et de préciser la spécification du besoin.

*3.2 Formalisme*

Notre formalisme doit donc présenter des concepts fonctionnels, tels que la notion de but ou d'agent, mais également des concepts techniques permettant de structurer la future application. Le nombre de concepts proposés doit être limité afin que la période d'apprentissage du formalisme soit la plus courte possible.

Notre proposition de formalisme s'articule autour de sept principaux concepts : *Entity*, *Relationship, Attribute*, *Organization*, *Agent*, *Goal* et *Privilege* (voir figure 3). Ils peuvent être répartis dans quatre grandes familles de modélisation :

- la modélisation de l'entreprise : elle a pour objectif de comprendre la structure dans laquelle l'application sera utilisée;
- la modélisation des objectifs : elle permet de décrire l'activité qui sera supportée par l'outil métier;
- la modélisation de la structure d'information : elle est utile afin de décrire les différents termes utilisés, c'est-à-dire le dictionnaire de données utilisé dans l'entreprise;

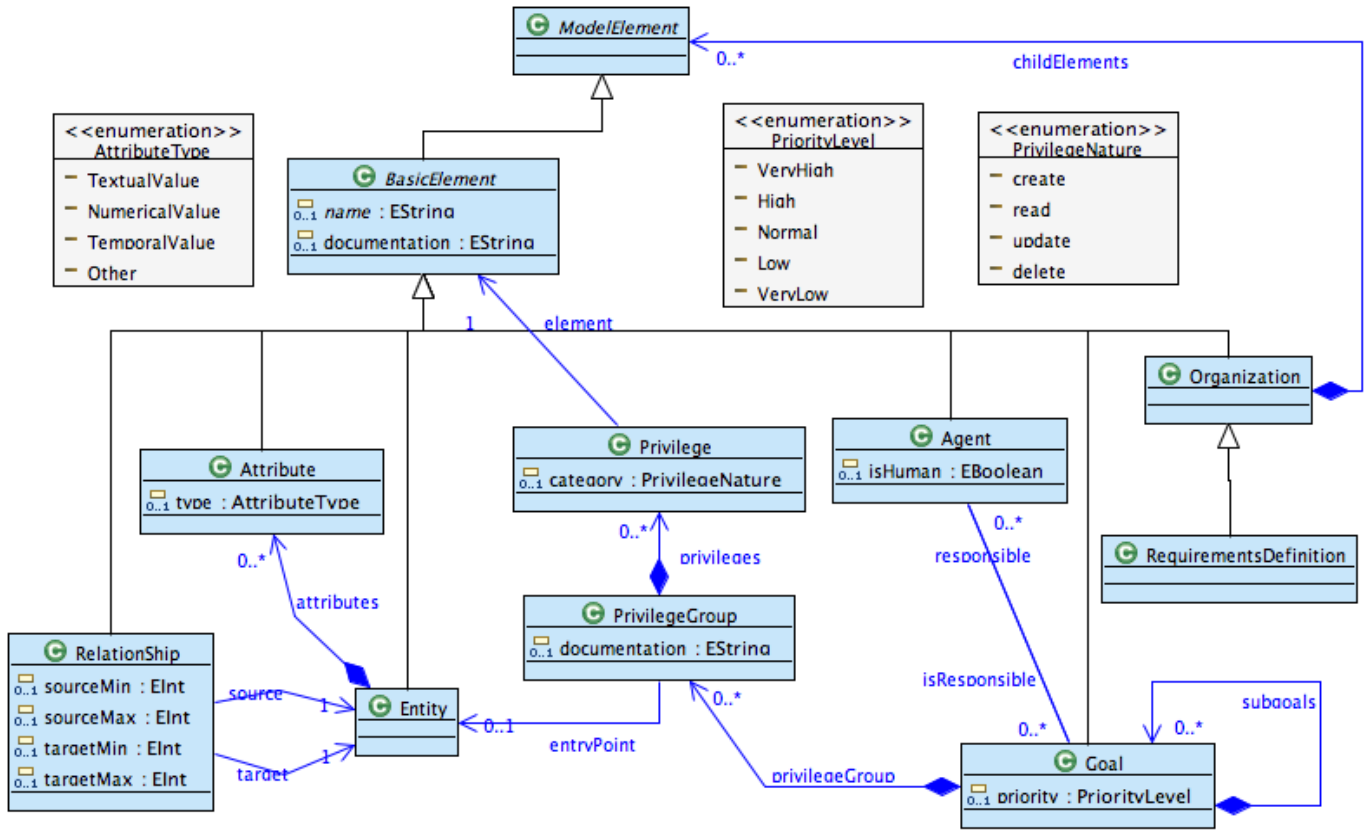

**Figure 3. Méta-modèle proposé pour l'expression des besoins**

- la modélisation des privilèges : elle permet de définir les moyens à mettre en oeuvre afin d'atteindre un objectif.

Ces différentes familles se concentrent sur différents aspects du futur outil métier et répondent à différentes questions utiles à la création du futur système. Cette décomposition permet ainsi d'obtenir un formalisme simple mais relativement complet. On peut classer celles-ci selon deux axes (voir figure 4). L'axe horizontal nous permet de les ordonner selon leurs capacités à répondre aux questions nécessaires à la création de l'outil. Il permet de répondre à la question : *Comment allons nous construire le futur outil métier ?* L'axe vertical nous permet de comprendre les raisons, les besoins et donc les activités supportées par le futur outil. Cet axe sera utile pour répondre à la question : *Pourquoi réalisons-nous cette application ?* Après avoir répondu à ces deux questions, nous avons donc compris les objectifs et limites de l'outil mais également comment le réaliser.

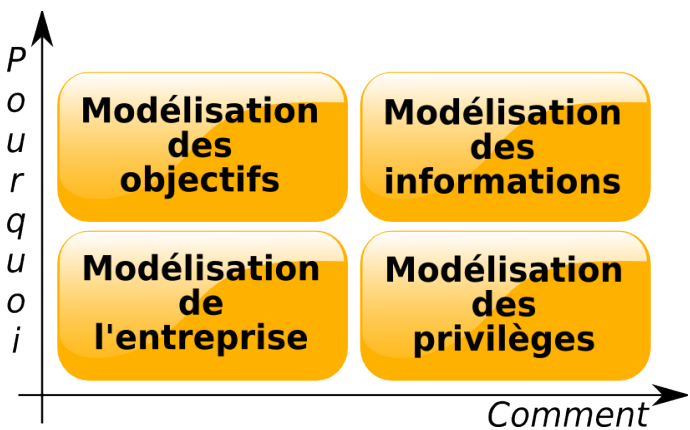

**Figure 4. Organisation des différentes modélisations**

Exemple :

On retrouve sur la figure 5 une expression partielle des besoins pour un système d'organisation de conférence. La partie à laquelle nous nous intéressons dans cet exemple est la soumission d'article. La première étape est de lire la partie concernant la modélisation des objectifs. L'objectif « Gérer les soumissions » se décompose en deux sous-objectifs. Tout d'abord, il y a la soumission avec l'objectif « Déposer une soumission » sous la responsabilité d'un auteur. Puis, il y a la re-lecture avec l'objectif « Analyser une soumission » sous la responsabilité d'un re-lecteur.

La seconde étape se concentre sur la modélisation des informations, c'est-à-dire les entités et relations entre elles ainsi que les attributs caractérisant les entités. Nous pouvons considérer la définition des concepts métier comme un graphe. Celui-ci est réduit dans cet exemple à deux noeuds représentés par les entités *article* et *rapport* reliés entre eux par une relation.

Après avoir défini ces deux modélisations, il est possible de les lier entre elles à l'aide des privilèges. Pour que les agents atteignent leurs buts dans le futur outil, nous allons leur définir un parcours dans le graphe des concepts, à l'aide des privilèges, pour chaque objectif sous leur responsabilité. Un parcours dans le graphe des concepts peut également être considéré comme une vue partielle sur le futur système d'information. La soumission d'un article nécessite un parcours assez simple. Il se résume à entrer dans le système par l'entité *article* puis de pouvoir en créer un nouveau. Ce parcours permet à l'auteur de renseigner le titre de l'article et ses auteurs. La re-lecture d'un article nécessite un parcours relativement plus complexe. Le point d'entrée dans le système est tout d'abord le *rapport*, un re-lecteur doit donc créer un nouveau rapport, saisir son commentaire mais aussi définir quel est l'article qui est commenté. Ce but doit donc permettre de naviguer jusqu'au concept d'*article* pour récupérer l'information nécessaire. Malheureusement, la figure 5 ne permet pas de visualiser les privilèges accordés pour un souci de clarté du diagramme. Le seul lien visible pour le concept de privilège est le point d'entrée dans le système d'information pour chacun des objectifs.

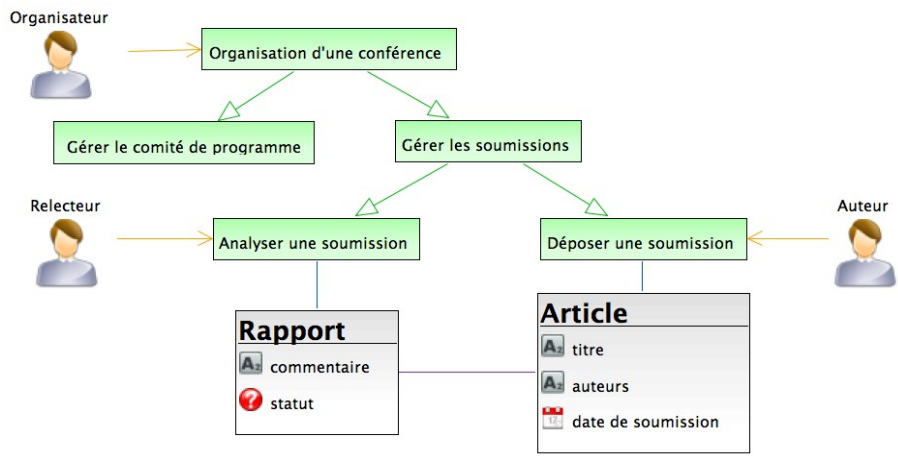

**Figure 5. Expression de besoins réalisée à l'aide de la proposition de formalisme**

*3.3 Interprétation automatique*

L'organisation traditionnelle présente donc des inconvénients. L'un d'eux est la trop grande sensibilité de l'étape d'interprétation par rapport aux conséquences sur la réussite du projet qui peuvent survenir en cas de spécification incomplète ou erronée. Afin de mieux répondre aux besoins du client et plus rapidement, il est conseillé, dans cette approche, de limiter cette étape d'interprétation manuelle.

Il existe deux grandes familles d'interprétation. Nous ne pouvons pas considérer un client comme un unique profil fonctionnel mais plutôt comme un ensemble de profils fonctionnels. L'expression des besoins va être conduite par un chargé de projet, un profil opérationnel (les futurs utilisateurs de l'application), un informaticien... Il est donc important d'adapter la visualisation de l'expression des besoins aux différents profils qui devront valider les spécifications fonctionnelles. Cette adaptation regroupe donc l'ensemble de transformations constituant la première famille. La seconde regroupe les transformations dites de « traduction ». Leur objectif est de lire l'expression des besoins, la comprendre et enfin la traduire sous diverses formes. On peut citer les transformations d'analyse, de vérification, de documentation ou de traduction vers un formalisme technique. En effet, à la fin du processus d'expression des besoins, il est nécessaire d'obtenir un ensemble d'informations compréhensibles par un expert système. Dans cet ensemble, nous y retrouverons des modèles techniques obtenus, en partie, à l'aide de déductions faites à partir de l'expression des besoins, et d'autre part, à l'aide de suppositions.

La première famille d'interprétation, c'est-à-dire celles qui vont permettre d'instaurer une meilleure communication et une meilleure compréhension des besoins par l'ensemble des participants du projet, regroupe les transformations vers les cartes conceptuelles[Buzan et al., 2006], les outils de validation fonctionnelle... Si on considère l'exemple des cartes conceptuelles, celles-ci sont généralement bien comprises et connues par les profils de type *manager*. Elles vont permettre de se focaliser sur un aspect de la spécification du besoin et/ou sur un fragment de cette même spécification. Elles vont ainsi nous permettre de zoomer sur la définition des concepts ou de sélectionner seulement les objectifs d'un agent spécifique dans la future application.

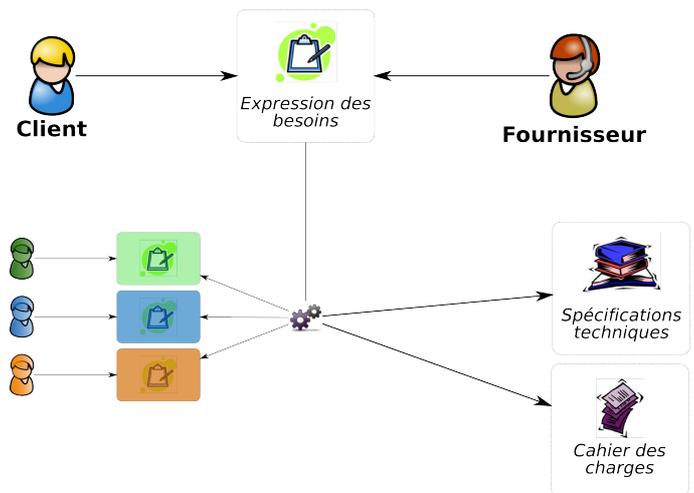

**Figure 6. Proposition d'organisation d'une communication entre un organisationnel et un profil technique**

La représentation sous forme de cartes conceptuelles peut devenir totalement inadéquate si la communication a lieu avec des profils opérationnels qui seront chargés d'utiliser le futur système. Ils seront plus à même de critiquer et de commenter la spécification des besoins si celle-ci est représentée sous la forme d'interface, à l'aide d'un prototype, en opposition à une représentation abstraite telle qu'une carte conceptuelle. Pour réaliser une interprétation de cette catégorie, nous pouvons réaliser une interprétation vers le méta-modèle WebML [Acerbis et al., 2008][Brambilla et al. 2008]. Celui-ci permet, à l'aide de l'outil WebRatio, de modéliser une application Web complexe puis de la générer. Cette modélisation est structurée selon trois axes : la définition de la structure de donnée, la navigation et la présentation. On peut alors réaliser une transformation de l'expression des besoins conforme à notre méta-modèle vers un modèle conforme à WebML exploitable par WebRatio pour ensuite générer l'application. Une autre interprétation possible est le diagnostic de l'expression du besoin. Ce diagnostic va nous permettre de vérifier des contraintes telles que :

- tous les agents sont responsables d'au moins un objectif,
- tous les agents sont responsables d'au moins un objectif,

- tous les objectifs permettent d'accéder au système d'information,
- …

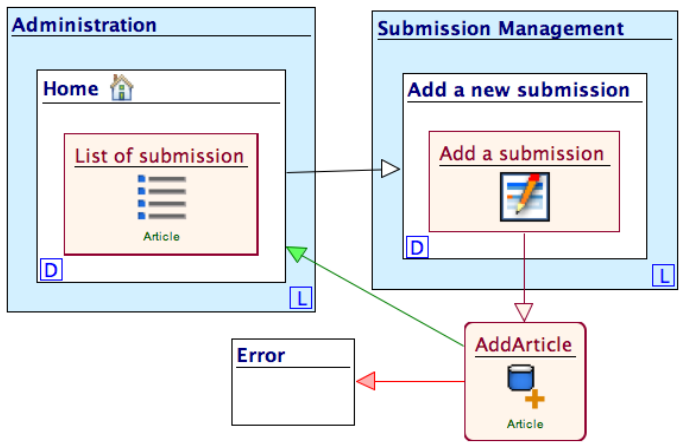

**Figure 7. Modèle d'application réalisé avec WebRatio**

La seconde famille d'interprétation regroupe les transformations permettant d'adapter la spécification du besoin dans un autre espace technologique puis d'utiliser le résultat pour initialiser un nouveau processus. Elle renferme donc les interprétations de documentation tel que la génération d'un cahier des charges qui pourra ensuite être complété. Elle renferme également des transformations vers des modèles d'application ré-utilisables par des outils présents sur le marché. Ces modèles d'application peuvent être conformes à WebML (voir figure 7), UML ou à des méta-modèles spécifiques (DSL) à l'outil cible.

## 4 Résultat

Un méta-modèle de spécification du besoin fonctionnel est aujourd'hui proposé. Celui-ci est supporté par un outil afin de manipuler ce formalisme et donc exprimer son besoin. Une première version de l'outil a été réalisé en mode Web, c'est-à-dire directement utilisable dans le navigateur. Cette possibilité permet de faciliter l'expression des besoins par des experts métier en simplifiant son accès. Le temps nécessaire à sa réalisation et à sa maintenance est très important en raison de l'immaturité des technologies utilisées dans le contexte de la création d'un modeleur. De plus, la compréhension des modèles avec cette première version est très difficile en raison de l'absence de fonctionnalités indispensables. De plus, la notion de modèle et de diagramme y sont fusionnés ce qui rend la modélisation et la transformation plus complexe. Nous avons donc choisi de réaliser une nouvelle version de l'outil intégré dans l'environnement de développement Eclipse. Celui-ci a été généré en suivant l'approche MDD/MDA proposé par le projet Topcased.

Différentes interprétations automatiques ont ete réalisées. Parmi celles-ci, on retrouve les cartes conceptuelles (ou *mind map* [Buzan et al., 2006]) qui représentent un moyen de communication simple et pratique pour les profils de type *manager*. Il a également été possible de transformer l'expression des besoins vers un document écrit et ainsi le diffuser plus facilement. Une autre transformation nous permet de diagnostiquer l'expression des besoins en vérifiant un ensemble de contraintes et pouvoir ainsi la corriger avant diffusion.

## 5 Conclusion

La première proposition du formalisme est aujourd'hui finalisée. Il permet d'exprimer son besoin pour ensuite l'interpréter correctement vers des représentations visuelles, textuelles ou techniques. Il est désormais important de le tester dans des projets industriels afin de vérifier son potentiel d'utilisabilité. Il est également important de faire valider le méta-modèle en le comparant à des formalismes existants de spécification du besoin sous forme d'objectif mais aussi à une description textuelle ou à un modèle UML. Pour cela, nous pensons utiliser les travaux de S. Patig [Patig, 2008] pour réaliser notre étude.

Une méthodologie doit être proposée pour utiliser pleinement le formalisme, l'outil et l'ensemble des interprétations. Cette méthodologie devra définir un guide pour spécifier quand, pour qui et comment les interprétations devront être utilisées. Elle permettra également de définir un processus traditionnel de capture et de spécification du besoin.

Les interprétations, mises à disposition aujourd'hui, nous ont permis de valider la méthodologie globale. De nombreuses autres interprétations sont en cours de réalisation. Nous nous intéressons particulièrement à l'analyse des besoins et à l'automatisation des différentes tâches du processus de développement. L'analyse des besoins nous permettrait d'isoler les objectifs ou les agents à risque potentiellement critique dans la future application. Elle regroupe également l'ensemble des interprétations de mesures quantitative et qualitative. En ce qui concerne l'automatisation du processus de développement, il serait envisageable de proposer des interprétations vers des documentations utilisateur, des cahiers de recette, des scénarios donc des interprétations plus techniques.

## 6 Références